# Fe Doped Magnetic Nanodiamonds Made by Ion Implantation as Contrast Agent for MRI

Bo-Rong Lin[1,2], Chien-Hsu Chen[2], Srinivasu Kunuku[2], Tzung-Yuang Chen[2],

Tung-Yuan Hsiao[2], Huan Niu[2]* and Chien-Ping Lee[1]

*(1) Institute of Electronics, National Chiao Tung University, Hsinchu, Taiwan*
*(2) Accelerator Laboratory, Nuclear Science and Technology Development Center,*
*National Tsing Hua University, Hsinchu, Taiwan*

**The use of nanodiamonds (NDs) for medical applications is a subject of intensive research in recent years[1]. The bio-compatibility and the versatility of these nano size particles have made them attractive for potential applications in drug delivery, disease diagnosis, and targeted therapy. The most attractive feature of NDs is their capability to be functionalized by surface modification. Biomolecules and other functional groups can be easily attached to their surfaces chemically enabling them to load drugs, to target specific proteins and DNA sequences[2, 3] performing specific tasks for different medical purposes. But most of the studies conducted so far are concentrated on the surfaces of NDs. People have not yet thought about how to modify the inside of these tiny particles for medical purposes. Although some of the defects or impurities introduced inside the NDs during the production processes can change some of the physical properties of NDs, they are difficult to be modified once they are produced because of the tight bonding of the carbon atoms. Recently we have developed a new technique using ion implantation to add impurities into the NDs. This technique allows us to alter the physical properties of NDs and adds a new dimension for their applications. Magnetic NDs have been produced by implanting Fe ions into the NDs. Excellent magnetic property has been demonstrated[4]. Using the magnetic properties of Fe doped NDs, we report in this paper the magnetic resonance imaging (MRI) capability of these NDs. Because the Fe atoms are embedded inside the NDs, the particles do not have cell toxicity. These particles can be monitored and used as an excellent contrast agent for MRI. The image enhancement properties of Fe doped NDs and the cell viability when they are used are reported for the first time. Combing the imaging capability of Fe doped NDs and their other functions, we foresee them a potential tool for performing drug delivery[5], radio frequency thermal therapy[6] and imaging at the same time.**

Magnetic resonance imaging (MRI) is a powerful tool in diagnosis of various diseases[7, 8]. More and more advanced functions and techniques are being developed to enhance the capability of MRI. To get precise diagnosis, one needs clear, high-resolution and high-contrast MR images. One of the keys to achieve such images is to have a good contrast agent. The contrast agents that are commonly used today are either gadolinium (Gd) based complex[9] or Fe based nanoparticles[10]. The magnetic properties of these materials enable them to enhance the MR images. However it has been shown that the Fe based contrast agents are toxic to living cells and therefore have been banned from clinical use for most of countries[11]. Today the most popular commercially available medical grade MRI contrast agents are Gd based complex. However they are not totally free of toxicity to human bodies[12]. Recently, the European Medicines Agency has decided to suspend the marketing authorization of some Gd based contrast agents used in body MRI scanning because of potential damages to the brain cells(7 July 2017 EMA/424715 /2017).

So what can we do? We need something that is both magnetic for MRI and at the same time nontoxic to human bodies. In this paper we demonstrate a nanoparticle based contrast agent that can do just that. Using the new technique we recently developed, we are able to dope NDs with Fe atoms using ion implantation. The accelerated Fe ions penetrate into the diamond particles and stay inside. We have shown in our recent study that these NDs possess excellent ferromagnetic property because of the presence of the Fe atoms[4]. Since NDs are known to be bio-compatible[13, 14, 15, 16] and the toxic Fe atoms are inside the particles with no direct contact with the outside world, they are safe to living cells and tissues that are subject to this magnetic nanoparticles. These magnetic nanoparticles are responsive to external magnetic fields so they can work effectively as a contrast agent for MRI.

Ion implantation is a common and popular technique used in the semiconductor industry. It puts desired impurity atoms into semiconductors to serve as active dopants providing the needed electrons and holes for device operations. It is proven that the atoms implanted into the semiconductors are stable and do not move in subsequent processes and during device operation. This point is very important since we intend to use this technique for medical purposes. The stability of the Fe atoms inside the NDs guarantees the safety of these nanoparticles for extended use in human bodies. NDs are being widely studied for medical applications[17]. They are not only nontoxic but also very versatile as they can be easily functionalized by attaching biomolecules to perform specific medical tasks, such as drug delivery[18], cell labeling[19] and targeted therapy[20]. In the past, there were also efforts trying to fabricate magnetic NDs for image enhancement and for localized thermal treatment. Gd-ND conjugates have been reported for MRI[21]. Magnetic NDs with Fe particles attached on the surface have also been reported[22]. But all these approaches used surface modification of the NDs. The toxicity associated with these magnetic materials can not be eliminated. The ion implantation process involves physical bombardment of Fe atoms onto the NDs. It is simple and straightforward and does not require any chemical processes.

The effectiveness of these magnetic NDs as a contrast agent was studied using a Bruker MRI scanner (BIOSPEC 70/30 MRI). The DC magnetic field used was 7.0 Tesla. T2 relaxation time was measured and the T2-weighted MR images were taken. The preparation procedure for the samples and other experimental parameters are described in the methods section. The 7.0 Tesla static magnetic field was first applied to polarize the spins of the water protons in the samples. As the net magnetization has been aligned with the static field, a radio frequency (RF) signal was applied to tip the magnetization towards the transverse direction. When the RF signal is removed, the net spin magnetization starts to relax back to the longitudinal direction. As the spins magnetization relax to their equilibrium state, the transverse component of the magnetization was monitored by a RF coil receiver. The transverse relaxation time constant, which is related to spin-spin interaction, is the so called T2.

Two sets of samples were prepared. One set contained NDs without Fe and the other Fe doped NDs. The NDs were dissolved in DI water. Six samples in each set were prepared with NDs at different concentrations, which varied from 0 to 2.4 *mg/ml*. The measured T2 values for the samples are listed in Table 1. For samples with regular NDs, T2 drops slightly as the concentration increases. But for samples with Fe doped NDs, the T2 values were significantly shortened and had a strong dependence on the concentration. The efficiency of contrast enhancement of our samples was calculated based on the 1/T2 versus concentration relationship. The rate of change or slope of this relationship gives the relaxivity, $r_2$, which is a measure of the contrast enhancement efficiency. Figure 1 shows the 1/T2 vs. concentration plots for the two sets of our samples. Very good straight line relationship was obtained. The calculated $r_2$ for the regular NDs was 0.145 $mls^{-1}mg^{-1}$ but for the Fe doped NDs, this value jumps to 0.951 $mls^{-1}mg^{-1}$. There is almost a seven times increase in $r_2$ for the Fe doped NDs.

The T2-weighted images of all samples are shown in Figure 2 (a). The images are obviously darker for samples with Fe doped NDs and the darkness intensifies as the concentration goes higher. The darkest image, the one with the lowest signal intensity, is clearly seen for the sample with the highest concentration of Fe doped NDs. We used the percentage of intensity enhancement defined as

$$\frac{I(Sample) - I(DI\ water)}{I(DI\ water)} \times 100\%$$

to quantify the image enhancement. These value are shown in the last column of Table 1 and the data are plotted against the NDs concentration shown in Figure 2(b). For the sample with a concentration of 2.4 *mg/ml* Fe doped NDs, there is a staggering 78.5% enhancement in signal intensity compared to the DI water control sample. This extra contrast enhancement is resulted from the Fe atoms inside the NDs. The Fe atoms generate local magnetic field inhomogeneities. Water protons near Fe-doped NDs become dephased by the local field inhomogeneities and T2 is shortened because of the extra spin-spin interactions. This causes the image to become darker when Fe doped

NDs are present in the sample.

The cytotoxicity of the Fe doped NDs has also been studied. Cell viability test was carried out by MTT assay and the living cells were the NCTC cloned 929 cell line. Figure 3 is the result of this cell viability test. It is seen clearly Fe doped NDs have very little effect on the living cells. The difference between those of the regular NDs and the Fe doped ones is insignificant. The micrographs of the cells incubated with Fe doped NDs are shown in Figure 4. We can see that the living cells can grow normally with the presence of Fe-doped NDs. The safety of Fe doped NDs demonstrated here is of great importance for them to be used as a contrast agent in human bodies. It shows that the Fe atoms implanted inside the NDs without direct contact with the living cells are safe to the cells. This is very different from the conventional Fe based contrast agent, known as super paramagnetic iron oxide (SPIO). Because the iron oxides are in direct contact with the living tissues and can easily be released to the cells they are in contact with, and therefore are not free from toxicity. With high T2 relaxivity and ultralow cell toxicity, Fe doped NDs have the potential to be MRI contrast agents and are safe for the following bio-applications.

In conclusion, we have demonstrated the capability of Fe doped magnetic NDs as contrast agents for MRI. The concentration dependent contrast enhancement and T2 were studied. The T2 relaxivity, $r_2$, is nearly seven times higher for Fe doped NDs compared with regular NDs. The Fe doping is obtained by ion implantation, which is simple and does not involve any chemical process. The cell viability test shows that the Fe atoms inside the NDs are safe for the living cells. The Fe doped magnetic NDs, when combined with other capabilities of NDs, are potentially useful for imaging, targeted cancer therapy, and localized treatment all at the same time.

**Methods.**

Fabrication of magnetic Fe doped magnetic nanodiamonds.

ND powder (average 100nm in diameter, Microdiamant Co.) was first dissolved in deionized water (DI water). The solution was then applied onto an oxidized silicon wafer and dried. The wafer was Fe-ion implanted with an energy 150 *keV* and dose $5 \times 10^{15}$ *atoms/cm²*. The methods to remove the NDs from the Silicon wafer and to collect the implanted magnetic NDs have been described in our previous work[4].

Relaxivity Measurement and MR imaging

Fe doped NDs were serially diluted with DI water to give five different sample concentrations (2.4, 1.2, 0.6, 0.3, and 0.15 *mg/ml*). Control samples were DI water and regular NDs with the same concentrations. All samples were contained in 0.3mL Eppendorf tubes, which were soaked in DI water to allow appropriate image acquisition. Bruker BIOSPEC 70/30 MRI scanner equipped with proper gradient coils was used. A multislice multiecho (MSME)-T2 map pulse sequence with fixed TR = 2700 *ms*, 60 echoes in 11 *ms* intervals, matrix size = 256 x 256, field of view = 60 x 60 *mm²*, slice thickness = 1 *mm*, an average of 1 was used to measure the spin-spin relaxation times(T2) and capture T2-weighted images. The inverse of the transversal relaxation time of each sample (1/T2, $s^{-1}$) was plotted against the concentration of Fe doped NDs and NDs. The relaxivity was extracted from the slope of linear regression curve. The shown T2-weighted image is the image from the 20th echo with TR = 2700 *ms* and TE = 649 *ms*.

MTT Assay using NCTC clone 929 cells

The cytotoxicity effect of NDs and Fe-NDs on NCTC clone 929 cells was evaluated by using a 3-(4,5-dimethylthiazol-2-yl)-2,5-diphenyltetrazolium bromide (MTT) assay. NDs and Fe-NDs samples were sterilized at 121℃ for 30 minutes. Before adding into cell culture, all samples were serial diluted with water and ultrasonic-treated for 30 minutes. NCTC clone 929 cells were seeded in a 96-well flat bottom plate as $1×10^4$ cell/well and incubated at 37℃ and 5% $CO_2$. After overnight culture, NDs or Fe-NDs was added into medium individually with designed concentrations 0.24 *mg/ml*, 0.06 *mg/ml*, 0.015 *mg/ml*, 0.00375 *mg/ml*, 0.00094 *mg/ml*, 0.00023 *mg/ml* and incubated for 24 h. Water was added as control. To evaluate cell survival, 10 *μl* of MTT solution was added to each well and incubated for an additional 4 hours. Then the media was replaced with 100 *μl* of DMSO and mixed thoroughly to yield soluble formazan. The absorbance was then determined at 540 *nm* by ELISA reader and compared to the control solution to measure the cytotoxicity effect. All conditions were performed in triplicate.

**Acknowledgements**

This work was financially supported by Team Union Ltd. The authors acknowledge Dr. Yu-Jen Chang and Ms. Li-Chuan Liao of the Food Industry Research and Development Institute(FIRDI) for the technical supports and valuable discussion.

**Author contributions**

H.N., C.H.C. and C.P.L. conceived of the project. H.N., C.H.C., B.R.L. and S.K. designed the ion implantation experiments, which were performed by B.R.L. and S.K. B.R.L. and S.K. designed the T2 and MRI experiments and B.R.L. performed those experiments. B.R.L. and T.Y.H. analyzed MRI data. T.Y.C. and B.R.L. designed the cell viability experiments, which were performed by FIRDI. B.R.L. wrote the paper with contributions from all authors. B.R.L. and C.H.C. have same contribution in this paper.

**Competing financial interests**

The authors declare no competing financial interests

**Materials & Correspondence**

Correspondence and requests for materials should be addressed to H.N. (email: hniu@mx.nthu.edu.tw)

**Tables**

Table 1. T2 measurement results and MRI intensity enhancement for ND and Fe-ND samples.

| Sample Name | Concentration(*mg/ml*) | T2(*ms*) | Enhancement(%) |
|---|---|---|---|
| ND | 0(DI Water) | 804.402±1.471 | 0 |
| | 0.15 | 785.833±1.484 | -8.064 |
| | 0.3 | 764.908±1.278 | -6.097 |
| | 0.6 | 741.179±1.390 | -5.391 |
| | 1.2 | 699.064±1.149 | -6.473 |
| | 2.4 | 623.116±1.137 | -14.544 |
| Fe-ND | 0(DI Water) | 804.402±1.471 | 0 |
| | 0.15 | 690.197±1.217 | -22.394 |
| | 0.3 | 662.526±1.423 | -23.266 |
| | 0.6 | 577.888±1.739 | -32.149 |
| | 1.2 | 428.878±3.604 | -53.063 |
| | 2.4 | 282.307±6.711 | -78.516 |

**Figures**

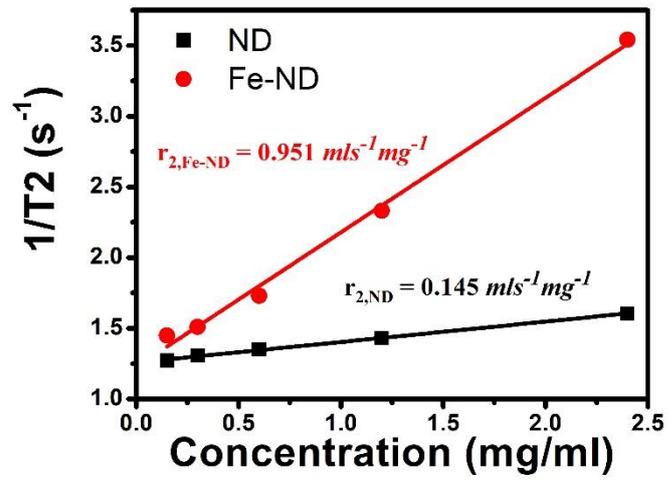

Figure 1. The inverse of the transversal relaxation time (1/T2, $s^{-1}$) against concentration plots of Fe-NDs and NDs samples.

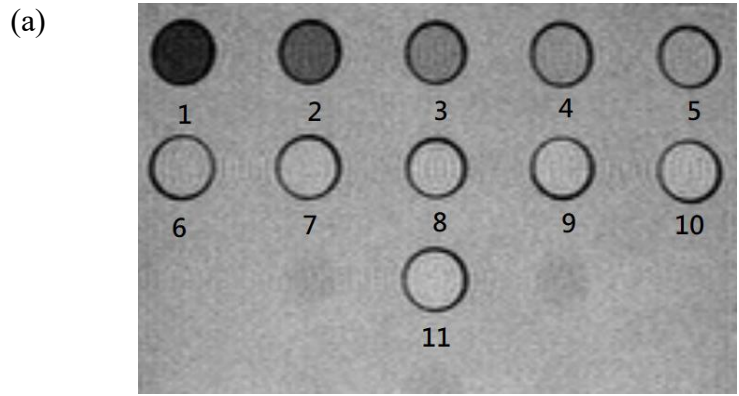

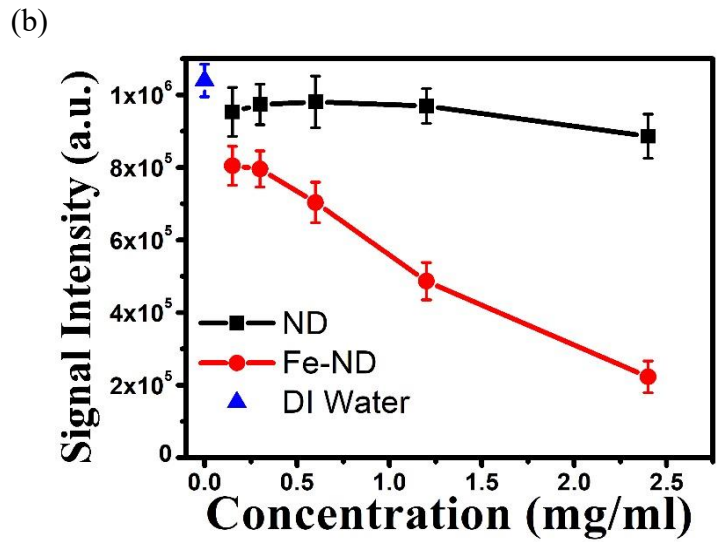

Figure 2. (a) T2-weighted image of Fe-NDs and NDs samples. 1, 2.4 *mg/ml* Fe-NDs; 2, 1.2 *mg/ml* Fe-NDs; 3, 0.6 *mg/ml* Fe-NDs; 4, 0.3 *mg/ml* Fe-NDs; 5, 0.15 *mg/ml* Fe-NDs; 6, 2.4 *mg/ml* NDs; 7, 1.2 *mg/ml* NDs; 8, 0.6 *mg/ml* NDs; 9, 0.3 *mg/ml* NDs; 10, 0.15 *mg/ml* NDs; 11, DI water. (b) MRI intensity plots of Fe-NDs and NDs samples. The intensity of DI water was used to be the intensity reference.

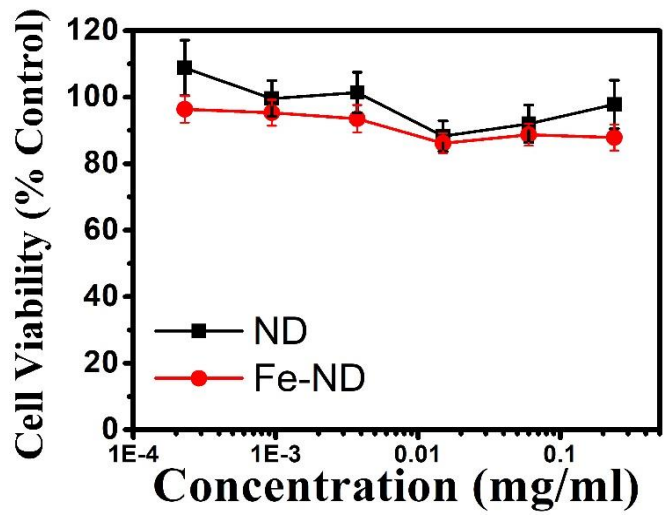

Figure 3. Cytotoxicity effects of Fe-NDs and NDs on NCTC clone 929 cells by MTT test. Cell viability is compared to control and represented as mean ± SD. All conditions were done in triplicate. No significant cytotoxicity effects were observed at concentrations up to 0.24 *mg/ml* in ND and Fe-ND treatment.

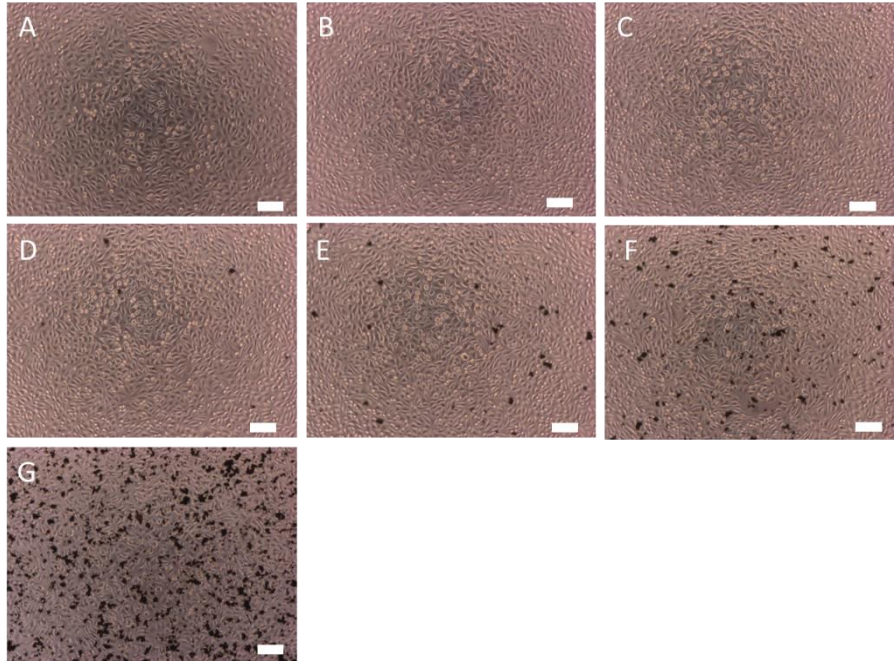

Figure 4. NCTC clone 929 Cell morphology under microscope after 24 hours Fe-ND treatment with different concentrations: (A) control (sterile water), (B) 0.00023 *mg/ml*, (C) 0.00094 *mg/ml*, (D) 0.00375 *mg/ml*, (E) 0.015 *mg/ml*, (F) 0.06 *mg/ml*, (G) 0.24 *mg/ml*. Scale bar: 100 *μm*.